%% file: arxiv.tex
\newcommand{\mykeywords}{galaxies: active -- galaxies: clusters:
  general -- X-rays: galaxies -- radio continuum: galaxies}
\newcommand{\samp}{21}
\newcommand{\phigh}{\ensuremath{P_{\rm{1.4}}}}
\newcommand{\pthree}{\ensuremath{P_{\rm{327}}}}
\newcommand{\plow}{\ensuremath{P_{\rm{200-400}}}}
\newcommand{\shigh}{\ensuremath{\sigma_{\rm{1.4}}}}
\newcommand{\sthree}{\ensuremath{\sigma_{\rm{327}}}}
\newcommand{\slow}{\ensuremath{\sigma_{\rm{200-400}}}}
\newcommand{\rhigh}{\ensuremath{r_{\rm{1.4}}}}
\newcommand{\rlow}{\ensuremath{r_{\rm{200-400}}}}
\newcommand{\mytitle}{A RELATIONSHIP BETWEEN AGN JET POWER AND RADIO POWER}
\newcommand{\mystitle}{AGN \pjet-\prad\ Relation}

\documentclass{emulateapj}
\usepackage{apjfonts,graphicx,here,common,longtable,ifthen,amsmath,amssymb,natbib}
\usepackage[pagebackref,
  pdftitle={\mytitle},
  pdfauthor={Dr. Kenneth W. Cavagnolo},
  pdfsubject={Astrophysical Journal Letters},
  pdfkeywords={\mykeywords},
  pdfproducer={LaTeX with hyperref},
  pdfcreator={LaTeX with hyperref}
  pdfdisplaydoctitle=true,
  colorlinks=true,
  citecolor=blue,
  linkcolor=blue,
  urlcolor=blue]{hyperref}
\begin{document}
\title{\mytitle}
\shorttitle{\mystitle}
\author{
K. W. Cavagnolo\altaffilmark{1,6},
B. R. McNamara\altaffilmark{1,2,3},
P. E. J. Nulsen\altaffilmark{3},\\
C. L. Carilli\altaffilmark{4},
C. Jones\altaffilmark{3},
and L. \birzan\altaffilmark{5},
}
\altaffiltext{1}{Department of Physics \& Astronomy, University of
  Waterloo, 200 University Ave. W., Waterloo, Ontario, N2L 3G1,
  Canada.}
\altaffiltext{2}{Perimeter Institute for Theoretical Physics, 31
  Caroline St. N., Waterloo, Ontario, N2L 2Y5, Canada.}
\altaffiltext{3}{Harvard-Smithsonian Center for Astrophysics, 60
  Garden St., Cambridge, Massachusetts, 02138-1516, United States.}
\altaffiltext{4}{National Radio Astronomy Observatory, P.O. Box 0,
  Socorro, NM 87801-0387, United States.}
\altaffiltext{5}{Leiden Observatory, University of Leiden, P.O. 9513,
  2300 RA Leiden, The Netherlands.}
\altaffiltext{6}{kcavagno@uwaterloo.ca}
\shortauthors{K. W. Cavagnolo et al.}
\slugcomment{Accepted to ApJ}


\begin{abstract}
  Using \chandra\ X-ray and VLA radio data, we investigate the scaling
  relationship between jet power, \pjet, and synchrotron luminosity,
  \prad. We expand the sample presented in \citet{birzan08} to lower
  radio power by incorporating measurements for \samp\ gEs to
  determine if the \citet{birzan08} \pjet-\prad\ scaling relations are
  continuous in form and scatter from giant elliptical galaxies (gEs)
  up to brightest cluster galaxies (BCGs). We find a mean scaling
  relation of $\pjet \approx 5.8 \times 10^{43} (\prad/10^{40})^{0.70}
  ~\lum$ which is continuous over $\sim 6-8$ decades in \pjet\ and
  \prad\ with a scatter of $\approx 0.7$ dex. Our mean scaling
  relationship is consistent with the model presented in \citet{w99}
  if the typical fraction of lobe energy in non-radiating particles to
  that in relativistic electrons is $\ga 100$. We identify several gEs
  whose radio luminosities are unusually large for their jet powers
  and have radio sources which extend well beyond the densest parts of
  their X-ray halos. We suggest that these radio sources are unusually
  luminous because they were unable to entrain appreciable amounts of
  gas.
\end{abstract} 


\keywords{\mykeywords}

\section{Introduction}
\label{sec:intro}

Most galaxies harbor a central supermassive black hole (SMBH) which
likely co-evolved with the host galaxy, giving rise to correlations
between bulge luminosity, stellar velocity dispersion, and central
black hole mass \citep{1995ARA&A..33..581K, magorrian}. Models suggest
these correlations were imprinted via galaxy mergers and the influence
of feedback from active galactic nuclei (AGN)
\citep[\eg][]{1998A&A...331L...1S, 2000MNRAS.311..576K}. Around the
time of these discoveries, the \cxo\ found direct evidence for AGN
feedback when observations revealed cavities and shock fronts in the
X-ray emitting gas surrounding many massive galaxies
\citep[\eg][]{perseus1, 2000ApJ...534L.135M}. X-ray cavities provide a
direct measurement of the mechanical energy released by AGN through
work done on the hot, gaseous halos surrounding them
\citep{2000ApJ...534L.135M}. This energy is expected to heat the gas
\citep{2001ApJ...554..261C} and prevent it from cooling and forming
stars.

Studies of X-ray cavities have shown that AGN feedback supplies enough
energy to regulate star formation and suppress cooling of the hot
halos of galaxies and clusters \citep{birzan04, 2005MNRAS.364.1343D,
  rafferty06}. Further supported by numerical simulations
\citep[\eg][]{croton06, bower06}, a consensus has emerged that AGN
feedback plays an important role in regulating galaxy evolution at
late times. However, the details of how AGN feedback is coupled to the
thermodynamics of the host cluster is still uncertain
\citep{2008ASPC..386..343D, 2009arXiv0910.3691M}. Studying feedback in
the broader cosmological context can be accomplished in principle
using radio observations to trace AGN activity
\citep[\eg][]{best07}. In order to do so, scaling relationships
between radio luminosity and mechanical jet power are required to
estimate \pjet\ in systems where X-ray observations of their halos are
either lacking or infeasible.

Relationships between \pjet\ and \prad\ were presented in
\citet[][hereafter B04]{birzan04} and \citet[][hereafter
  B08]{birzan08}. In B08, scaling relations between \pjet\ and 327
MHz, 1.4 GHz, and bolometric radio luminosities were discussed. B08
found $\pjet \propto \prad^{0.5-0.7}$ depending on the observed radio
frequency. However, there are few objects in the B08 study with $\prad
\lesssim 10^{38} ~\lum$ and $\pjet \lesssim 10^{43} ~\lum$, and the
relations have rather large scatter (discussed in Section
\ref{sec:relation}). In this paper we extend the study of B08 with the
inclusion of \samp\ gEs from systems with lower X-ray luminosities and
jet powers compared to those found in rich clusters.

We outline the sample of gEs in \S\ref{sec:sample}. X-ray and radio
measurements are discussed in \S\ref{sec:data}. Results and discussion
are presented in \S\ref{sec:r&d}. The summary and concluding remarks
are given in \S\ref{sec:summary}. \LCDM. All quoted uncertainties are
68\% confidence.

\section{Sample}
\label{sec:sample}

The \samp\ gEs in this study (see Table \ref{tab:sample}) were taken
from the sample of 160 gEs compiled by Jones et al. \citetext{in
  preparation}. The Jones et al. compilation is drawn from the samples
of \citet{1999MNRAS.302..209B} and \citet{2003MNRAS.340.1375O} using
the criteria that the $K$-band luminosity exceeds $10^{10}~\lsol$ and
the object has been observed with \chandra. Of the 160 gEs, AGN
activity was suspected in \samp\ objects (see Section \ref{sec:xray}
for details). Most of the gEs studied here have X-ray halos and radio
sources with luminosities lower than are typically found for cDs and
BCGs. The B08 sample is taken from \citet[][hereafter
  R06]{rafferty06}.

\section{Observations and Data Analysis}
\label{sec:data}

\subsection{X-ray}
\label{sec:xray}

Jet powers were determined in the usual manner from the X-ray data
(see B04 and R06). Gas properties used here are taken from the
analysis by Jones et al. (in preparation). Cavity locations and sizes
are from Nulsen et al. (in preparation), with cavity volumes and their
errors calculated by the method of B04. Cavities were identified based
primarily on the presence of surface brightness depressions in the
X-ray emitting gas and their association with radio emission, though
the latter is not requisite. To reflect this fact, each cavity system
is given a figure of merit (see Section \ref{sec:relation}). The
energy of each cavity was estimated as $4pV$, the enthalpy of a cavity
filled with relativistic gas. To estimate average cavity power, \pcav,
the energy of each cavity was divided by an estimate of its age. For
compatibility with B08, we give results derived assuming the age of
the cavities is approximated by the buoyant rise time, $t_{\rm{buoy}}$
\citep{2001ApJ...554..261C, 2003ApJ...592..839B}. Because the
mechanical power estimated from the cavities greatly exceeds the
synchrotron power of the radio source, we assume that $\pcav =
\pjet$. Further, \pcav\ does not include energy channeled into shocks
\citep[\eg][]{2007ApJ...665.1057F, 2009ApJ...707.1034B}, which can
exceed the energy in cavities \citep[\eg][]{herca}. The cavity power
estimates were summed for all the cavities in each system to obtain an
estimate of the average AGN outburst power. Because the uncertainties
are large (chiefly due to the uncertain volume estimates), errors are
propagated in log space, assuming that the uncertainties in ambient
pressure, cavity volume, and cavity age are independent of one
another.

\subsection{Radio}
\label{sec:radio}

Radio powers, $P_{\nu_0}$, were estimated using the relation
$P_{\nu_0} = 4 \pi D_L^2 (1+z)^{\alpha-1} S_{\nu_0} \nu_0$, where
$S_{\nu_0}$ is the flux density at the observed frequency, $\nu_0$,
over the integrated area of the source, $z$ is redshift, $D_L$ is
luminosity distance, and $\alpha$ is radio spectral index. The
redshift correction to alpha is small for our sample of nearby objects
and has been ignored. When no spectral index was available, we assumed
the spectra behave as $S_{\nu} \propto \nu^{-\alpha}$ with a spectral
index of $\alpha = 0.8$, which is typical for extragalactic radio
galaxies \citep{1992ARA&A..30..575C}. The 1.4 GHz radio flux for each
source was taken from the NRAO VLA Sky Survey (NVSS,
\citealt{nvss}). NGC 1553 lies outside the NVSS survey area, so the
1.4 GHz flux was estimated using the 843 MHz Sydney University
Molonglo Sky Survey (SUMSS, \citealt{sumss1}) flux and $\alpha =
0.89$; $\alpha$ was derived using the SUMSS and 5 GHz Parkes
\citep{1970ApL.....5...29W} fluxes.

The radio morphologies for our sample are heterogeneous: some are
large and extended, while others are compact. As a result, most
compact sources have a single catalog entry, while large sources are
divided among multiple entries. To ensure the entire radio source was
measured, a fixed physical aperture of 1 Mpc was searched around the
X-ray centroid of each gE. For each target field, all detected radio
sources were overlaid on a composite image of X-ray, optical (DSS
I/II), and infrared emission (2MASS). When available, the deeper and
higher resolution radio data from VLA FIRST was included. A visual
inspection was performed to establish which radio sources were
associated with the target gE. After confirming which catalog sources
are associated with the target gE, the fluxes of the individual
catalog sources were added and the associated uncertainties summed in
quadrature.

Archival VLA data for each source was also reduced and analyzed. The
continuum VLA data were reduced using a customized version of the NRAO
VLA Archive Survey reduction pipeline. In the cases where
high-resolution VLA archival data is available, multifrequency images
were used to confirm the connection between NVSS detections and the
host gE. Images at 1.4 GHz were further used to check NVSS fluxes. We
found flux agreement for most sources, the exceptions being IC 4296
and NGC 4782, where the NVSS flux is approximately a factor of 2
lower. The radio lobes for IC 4296 and NGC 4782 contain significant
power in diffuse, extended emission which is not detected in NVSS
because the NVSS flux limit is higher than the archival observations
used. For these sources, the fluxes measured from the archival VLA
data are used in our analysis. For the systems where nuclear radio
emission was resolved from lobe emission, we found
$S_{\nu_0,\rm{nucleus}} / S_{\nu_0,\rm{lobe}} < 0.2$, suggesting the
nuclear contribution to the low-resolution NVSS measurements has a
small impact on our results. NGC 6269 appeared to be the exception
with $S_{\nu_0,\rm{nucleus}} / S_{\nu_0,\rm{lobe}} \approx 1$,
however, this is based upon the NVSS data only.

B08 found that using lower frequency radio data, \ie\ 327 MHz versus
1400 MHz, resulted in a lower scatter \pjet-\prad\ relation. We
therefore decided to test this using our sample of gEs. Unfortunately,
the quality and availability of 327 MHz data for our gE sample were
not ideal, thus we gathered low-frequency radio fluxes from the CATS
Database \citep{cats}. The CATS Database is a compilation of more than
350 radio catalogs (\eg\ WENSS, WISH, TXS, B3). For each gE, the CATS
database was searched in the frequency range 200--400 MHz for a
counterpart to the NVSS and SUMSS sources. Of the \samp\ gEs in our
sample, 17 of them were found to have a radio source in the CATS
database. CATS does not provide images for visual inspection and is
composed of catalogs having a variety of spatial resolutions and flux
limits. Thus, the 200-400 MHz radio powers shown in Figure
\ref{fig:pcav} may include some contribution from background sources.

\section{Results and Discussion}
\label{sec:r&d}

\subsection{\pjet-\prad\ Scaling Relations}
\label{sec:relation}

The results from the X-ray and radio data analysis are shown in the
plots of \pcav-\phigh\ and \pcav-\plow\ presented in Figure
\ref{fig:pcav}. A figure of merit (FM) was assigned to each set of
cavities through visual inspection, shown with color coding in Figure
\ref{fig:pcav}. To give context to each FM, we supply a well-known
cluster system as an example. FM-A cavities have well-defined
boundaries and are coincident with radio emission which can be traced
back to an AGN (\eg\ Perseus); FM-B cavities are coincident with radio
emission from an AGN but lack well-defined boundaries (\eg\ A2597);
FM-C cavities have poorly-defined boundaries and their connection to
AGN radio activity is unclear (\eg\ A1795). FM-C cavities are excluded
from all fitting, as are a subset of objects we have defined as being
poorly confined (discussed in Section \ref{sec:jet} and excluded from
Figure \ref{fig:pcav}).

Figure \ref{fig:pcav} shows a continuous, power-law relationship
between cavity power and radio power spanning 8 orders of magnitude in
radio power and 6 orders of magnitude in cavity power. To determine
the form of the power-law relation, we performed linear fits in
log-space for each frequency regime using the bivariate correlated
error and intrinsic scatter (\bces) algorithm \citep{bces}. The
orthogonal \bces\ algorithm takes in asymmetric uncertainties for both
variables, assumes the presence of intrinsic scatter, and performs a
linear least-squares regression which minimizes the squared orthogonal
distance to the best-fit relation. Parameter uncertainties were
calculated using 10,000 Monte Carlo bootstrap resampling trials. Our
fits differ from the method used in B08 which minimized the distance
in the \pcav\ coordinate.

The best-fit log-space orthogonal \bces\ relations are:
\begin{eqnarray}
  \log~\pcav &=& 0.75~(\pm 0.14)~\log~\phigh + 1.91~(\pm 0.18) \label{eqn:high}\\
  \log~\pcav &=& 0.64~(\pm 0.09)~\log~\plow + 1.54~(\pm 0.12) \label{eqn:low}
\end{eqnarray}
where \pcav\ is in units $10^{42} ~\lum$, and \phigh\ and \plow\ are
in units $10^{40} ~\lum$. The scatter for each relation is $\shigh =
0.78$ dex and $\slow = 0.61$ dex, and the respective correlation
coefficients are \rhigh\ = 0.72 and \rlow\ = 0.81. We have quantified
the total scatter about the best-fit relation using a weighted
estimate of the orthogonal distances to the best-fit line
\citep[see][]{2009A&A...498..361P}. For comparison, the B08 scaling
relations are
\begin{eqnarray}
  \log~\pcav &=& 0.35~(\pm 0.07)~\log~\phigh + 1.85~(\pm 0.10) \\
  \log~\pcav &=& 0.51~(\pm 0.07)~\log~\pthree + 1.51~(\pm 0.12) \label{eqn:err}
\end{eqnarray}
where \pcav\ is in units $10^{42} ~\lum$, and \phigh\ and \pthree\ are
in units $10^{24}$ W Hz$^{-1}$ (or $\approx 10^{40} ~\lum$). The B08
relations have scatters of $\shigh = 0.85$ dex and $\sthree = 0.81$
dex. Equation 15 of B08 contains an error \citep{birzan08err}, and the
correct version is given in Equation \ref{eqn:err} above.

In contrast to B08, the slopes of the relations in this work now agree
to within their uncertainties. Note that we find a steeper
relationship at 1.4 GHz than B08. The difference in slope at 1.4 GHz
between our work and B08 is due to the additional data points at lower
\pjet\ and the different fitting method. The B08 points tend to be
clumped in a fairly narrow power range, which gave the few points at
the upper and lower power extremes excessive leverage over the
slope. The new data extends to lower jet powers, giving a more uniform
sampling and improved measurements of the slope and zero point.

The scatter in these relations is large, and B08 showed that
correcting for the effect of radio aging by including a scaling with
break frequency (\ie\ source age) reduces the scatter by $\approx
50\%$. The substantial scatter in the \pjet-\prad\ relations suggest
radio lobe properties which affect synchrotron emission, \eg\ age,
composition, or magnetic field configuration, make synchrotron
emission a poor surrogate for total jet power. Though we measure
similar scatters and slopes for the gEs as the clusters, the scatter
in \pjet-\prad\ may be particularly important for gEs, which have
steeper pressure profiles and are more susceptible to disruption by
AGN outbursts \citep{2006MNRAS.372.1161W, 2008ApJ...687L..53P}. We
discuss this issue further in Section \ref{sec:jet}.

\subsection{Comparison with Models and Observational Studies}
\label{sec:models}

Relations presented in \citet[][hereafter W99]{w99} are commonly used
to estimate total AGN kinetic power from observed radio power. It is
therefore useful to compare our results with those of W99. For
simplicity, in this section we use the parameterization $\pjet = \eta
\prad^{\Gamma}$, where \pjet\ is total kinetic jet power, $\eta$ is
some normalization, $\Gamma$ is a scaling index, and \prad\ is
emergent synchrotron power.

W99 derive $\Gamma$ and $\eta$ using the jet model of
\citet{1991MNRAS.250..581F} and assuming the radio lobes are at
minimum energy density \citep[see][for
  details]{1980ARA&A..18..165M}. W99 derived $\Gamma = 6/7 ~(\approx
0.86)$ with $\eta \approx f^{3/2}~4.61 \times 10^{41} ~\lum$ when
\prad\ is in units of $10^{40} ~\lum$. We have adjusted the fiducial
W99 normalization from 151 MHz to 1.4 GHz assuming $S_{\nu} \propto
\nu^{-0.8}$. The factor $f$ consolidates a variety of unknowns (see
W99 for details). The fiducial W99 model ($f=1$) yields $\eta$ two
orders of magnitude below our normalizations, but the slopes formally
agree (see Figure \ref{fig:radeff}).

The W99 normalization has a weak dependence on ambient gas density.
Using shallower and lower density gas profiles that are better matched
to X-ray observations than the profiles adopted by W99, faster jet
outflow velocities are found. Similarly, the fractional deviation from
the minimum-energy condition is assumed to be small. More importantly,
the W99 model depends strongly on $k$, which is the ratio of energy in
non-radiating particles to relativistic electrons. We find that for
$k$ lying in the range of tens to thousands, values consistent with
observational findings \citep{2005MNRAS.364.1343D,
  2006MNRAS.372.1741D, 2006ApJ...648..200D, birzan08}, the W99
normalization is brought into agreement with our work. The scatter in
our relations may arise from intrinsic differences in radio sources
(light and heavy jets), or because confined jets are born light and
become heavy on large scales due to entrainment. These results suggest
that typical lobes created by AGN jets contain a small fraction of
their energy in relativistic electrons. Many systems showing X-ray
cavities have relatively low radio powers.  Systems in which a greater
proportion of the energy is contained in relativistic electrons are
likely to be more radio luminous and so may tend to be overrepresented
in radio surveys.

For flat-spectrum, compact radio cores (\ie\ small scale jets and not
radio lobes), several jet models predict $\Gamma = 12/17 ~(\approx
0.71)$ \citep{1979ApJ...232...34B, 1995A&A...293..665F,
  2003MNRAS.343L..59H}. Observational studies by
\citet{2005ApJ...633..384H} and \citet{2007MNRAS.381..589M}, which
used nuclear radio powers to estimate \pjet, found $\Gamma$'s and
$\eta$'s consistent with our relations. The similarity of our
\pjet-\prad\ relations with these studies may be coincidental given
that our measurements are for integrated radio emission and not just
nuclear radio emission. They agree, perhaps, because nuclear flux and
total radio luminosity are correlated in these systems.

\subsection{Poorly Confined Sources}
\label{sec:jet}

The X-ray data are too shallow to image the extent of the cavities for
IC 4296, NGC 315, NGC 4261, NGC 4782, and NGC 7626. Their radio lobes
extend beyond the observed X-ray halo and the radio morphologies are
distinctly different from the rest of the gE and B08 objects. X-ray
cavities typically enclose the radio emission, as is seen, for
example, in M84 \citep{2001ApJ...547L.107F, 2008ApJ...686..911F} where
the interaction takes on some complexity (see Figure
\ref{fig:pics}). However, the radio sources appear to be breaking
through the X-ray halos of the objects listed above (see Figure
\ref{fig:pics} for an example in NGC 4261), suggesting that they may
be poorly confined (PC) by their X-ray halos. While the lobes do not
appear to be confined by the X-ray emitting gas, the following
analysis assumes that they do remain confined by lower pressure
gas. Note that the PC systems have been excluded from analysis of the
\pjet-\prad\ relations.

To compare the properties of PC sources with the rest of our sample,
we calculated \pcav\ values for these systems assuming the volume of a
cavity equals the volume of the corresponding radio lobe. Pressure
profiles were extrapolated to large radii using $\beta$-models
\citep{betamodel} fitted to the surface brightness profiles. We
assumed isothermal atmospheres and a background gas pressure of
$10^{-13}$ erg \pcc. The assumed background pressure is based on the
mean value observed in the outskirts of clusters and groups
\citep{accept}. Cavity buoyancy ages cannot be directly calculated
from the data. To maintain consistency with the rest of the study,
$t_{\rm{buoy}}$ was estimated by scaling the gas sound speed by 0.65
which is the mean value of the ratio $t_{\rm{cs}}/t_{\rm{buoy}}$ for
the B04 sample. As a result of the radio lobes extending into regions
where the pressure profiles are steep and approaching the background
pressure, the large lobe volumes are offset by low pressures and long
ages, resulting in modest values of \pcav\ relative to their radio
luminosities.

Two of the PC sources, NGC 315 and NGC 4261, are in a sample of nine
FR-I objects analyzed by \citet[][hereafter C08]{2008MNRAS.386.1709C}
using \xmm\ X-ray observations. C08 provides the $4pV$ cavity energy
and the mean temperature of the lobe environments for each FR-I
source. As with the PC sources, we calculated \pcav\ for each of the
C08 FR-I sources using scaled sound speed. For N315 and N4261 we find
no significant difference between \pcav\ calculated using the
\chandra\ data and \xmm\ data. \phigh\ was calculated for each C08
source using the method outlined in Section \ref{sec:radio}. The PC
and C08 objects are plotted in Figure \ref{fig:radeff}.

In Figure \ref{fig:radeff} we highlight the location of the PC sources
relative to our best-fit, 1.4 GHz relation and the fiducial W99
relation. Figure \ref{fig:radeff} shows that the PC and C08 FR-I
sources reside well below our best-fit relation. This discrepancy
implies that these sources have the lowest jet power per unit radio
power of all objects in the sample. One possible explanation is that
these sources have lower $k$ values than the rest of the sample. This
may arise naturally from matter entrainment along the jets, resulting
in higher $k$ values only when a more extended, dense halo is present
as is found in clusters. The radio sources in PC systems extend beyond
their bright halos, and thus might not entrain significant quantities
of matter. Were these sources located in denser cluster environments,
they would presumably have lower radio power because a larger fraction
of their energy flux would be carried in non-radiating
particles. Another explanation is that PC sources are systematically
more powerful than our measurements indicate due to energy being
imparted to shocks. On average, shock energy is a modest correction to
\pcav, factor of a few in clusters, but the incidence and variety of
AGN driven shocks is broad \citep[\eg][]{2003ApJ...592..129K,
  hydraa}. However, a shock explanation would require that the
fraction of jet power going into shocks is preferentially higher for
systems with relatively high implied radiative efficiencies.

\section{Summary and Conclusions}
\label{sec:summary}

We have presented analysis of the jet power versus radio power scaling
relation for the B08 sample and a sample of \samp\ giant elliptical
galaxies observed with the \cxo. Cavity powers were calculated for
each set of cavities using similar methods to those outlined in
R06. Radio powers for our sample were estimated using 1.4 GHz and
200-400 MHz fluxes taken from the NVSS/SUMSS surveys and the CATS
database, respectively. We find a continuous power-law relation
between \pjet\ and \prad\ covering 6 decades in \prad\ and 8 decades
in \pjet\ (Figure \ref{fig:pcav}). We find the power laws describing
the \pjet-\prad\ trend have the mean form $\pjet \approx 5.8 \times
10^{43} (\prad/10^{40})^{0.70} ~\lum$, and a scatter about the fit of
$\approx 0.7$ dex. Our relations agree reasonably well with previous
observational studies and they are consistent with theoretical
expectations if the fraction of the energy in relativistic electrons
is small in a typical radio lobe.

Several groups have applied the \birzan\ scaling relations to study
the effects of AGN feedback on structure formation,
\eg\ \citet{best07} and \citet{2007MNRAS.379..260M}, with some groups
now suggesting that distributed low-power radio galaxies may dominate
heating of the intracluster medium,
\eg\ \citet{2009ApJ...705..854H}. Up to now, the available
observational results, which were primarily calibrated to high-power
radio sources, did not clearly indicate if a \pjet-\prad\ relation
would be continuous, or of comparable scatter, for lower power radio
sources. Assuming there is no redshift evolution of \pjet-\prad, our
relations suggest higher mass galaxies dominate over lower mass
galaxies in the process of mechanical heating within clusters and
groups.

The subset of objects with radio sources that are poorly confined by
their hot halo have \prad/\pjet\ ratios which are large relative to
the rest of our sample (see Figure \ref{fig:radeff}). In addition, PC
sources reside in the same region of the \pjet-\prad\ plane as the
FR-I sources taken from C08. Radio emission from lobes depends on
their composition, so that the large scatter in the
\pjet-\prad\ relationship may result from processes such as gas
entrainment and shocks. It seems likely that some of the scatter
arises as radio sources age, but it remains unclear what other factors
are important. With better constraints on the relationships between
jet power and radio synchrotron power, our understanding of the
composition and mechanical power of radio sources has improved
considerably. Models and simulations of AGN feedback that successfully
reproduce these results will yield new insight into jet formation
mechanisms and the structure of the central engine.

\acknowledgements

KWC and BRM acknowledge generous support from the Natural Sciences and
Engineering Research Council of Canada and grants from the \cxo. CJ
thanks the Smithsonian Institution for generous support. PN thanks the
\cxo\ Center for supporting this work. KWC thanks Judith Croston,
David Rafferty, Lorant Sjouwerman, and Chris Willott for helpful
discussions. We also thank the anonymous referee for a prompt and
helpful review. The \cxo\ Center is operated by the Smithsonian
Astrophysical Observatory for and on behalf of NASA under contract
NAS8-03060. The National Radio Astronomy Observatory is a facility of
the National Science Foundation operated under cooperative agreement
by Associated Universities, Inc.


{\it Facilities:} \facility{CXO (ACIS)} \facility{VLA}


\bibliography{cavagnolo}


\clearpage
\input{sample.tex}
\clearpage
\input{figs.tex}

\end{document}

%% file: sample.tex
\begin{deluxetable}{lccccccccc}
\tablewidth{0pt}
\tabletypesize{\scriptsize}
\tablecaption{Summary of Sample\label{tab:sample}}
\tablehead{\colhead{Source} & \colhead{R.A.} & \colhead{Dec.} & \colhead{$D_L$} & \colhead{\phigh} & \colhead{\plow} & \colhead{$\pcav^{\mathrm{buoy}}$} & \colhead{$L_{\mathrm{X}}(<r_{\mathrm{cool}})$} & \colhead{FM} & \colhead{Ref.}\\
\colhead{-} & \colhead{h:m:s} & \colhead{$\mydeg:\arcm:\arcs$} & \colhead{Mpc} & \colhead{$10^{39}~\lum$} & \colhead{$10^{39}~\lum$} & \colhead{$10^{42}~\lum$} & \colhead{$10^{33}~\lum$} & \colhead{--} & \colhead{--}\\
\colhead{(1)} & \colhead{(2)} & \colhead{(3)} & \colhead{(4)} & \colhead{(5)} & \colhead{(6)} & \colhead{(7)} & \colhead{(8)} & \colhead{(9)} & \colhead{(10)}}
\startdata
\dataset [ADS/Sa.CXO#Obs/03394] {IC 4296}$^{\dagger}$  & 13:36:39 & -33:57:57 & 54.3 & $38.9   \pm 3.5$    & $190    \pm 12$     & $3.87^{+1.44}_{-3.03}$ & 18.6  & B & [1]\\
\dataset [ADS/Sa.CXO#Obs/04053] {NGC 193}              & 00:39:19 & +03:19:52 & 63.5 & $8.60   \pm 0.27$   & $7.98   \pm 0.40$   & $9.99^{+3.12}_{-4.54}$ & 45.7  & A & [2]\\
\dataset [ADS/Sa.CXO#Obs/00855] {NGC 315}$^{\dagger}$  & 00:57:49 & +30:21:09 & 73.1 & $59.2   \pm 5.4$    & $21.5   \pm 1.5$    & $6.58^{+2.48}_{-4.96}$ & 28.8  & B & [3]\\
\dataset [ADS/Sa.CXO#Obs/00317] {NGC 507}              & 01:23:40 & +33:15:20 & 83.9 & $1.16   \pm 0.05$   & $1.85   \pm 0.19$   & $19.9^{+4.4}_{-6.8}$   & 1099  & B & [4]\\
\dataset [ADS/Sa.CXO#Obs/05001] {NGC 777}              & 02:00:15 & +31:25:46 & 63.6 & $0.0473 \pm 0.0034$ & $0.0376 \pm 0.0038$ & $4.08^{+1.48}_{-3.03}$ & 98.5  & C & [5]\\
\dataset [ADS/Sa.CXO#Obs/02022] {NGC 1316}             & 03:22:42 & -37:12:30 & 21.6 & $0.198  \pm 0.008$  & $0.0476 \pm 0.0081$ & $1.11^{+0.26}_{-0.40}$ & 5.93  & B & [6]\\
\dataset [ADS/Sa.CXO#Obs/00783] {NGC 1553}             & 04:16:11 & -55:46:49 & 18.6 & $0.0039 \pm 0.0006$ & $0.202  \pm 0.067$  & $0.98^{+0.34}_{-0.70}$ & 172   & C & [7]\\
\dataset [ADS/Sa.CXO#Obs/04283] {NGC 1600}             & 04:31:40 & -05:05:10 & 58.0 & $0.346  \pm 0.015$  & \nodata             & $1.87^{+0.82}_{-2.00}$ & 11.6  & B & [8]\\
\dataset [ADS/Sa.CXO#Obs/02073] {NGC 3608}             & 11:16:59 & +18:08:55 & 23.0 & $< 0.0022$          & $0.0336 \pm 0.0002$ & $0.05^{+0.02}_{-0.03}$ & 0.783 & C & [9]\\
\dataset [ADS/Sa.CXO#Obs/09569] {NGC 4261}$^{\dagger}$ & 12:19:23 & +05:49:31 & 31.8 & $17.7   \pm 0.5$    & $61.4   \pm 6.5$    & $0.91^{+0.37}_{-0.79}$ & 7.57  & B & [10]\\
\dataset [ADS/Sa.CXO#Obs/06131] {NGC 4374}             & 12:25:04 & +12:53:13 & 18.5 & $3.47   \pm 0.12$   & $2.72   \pm 0.28$   & $5.03^{+2.17}_{-5.53}$ & 7.58  & A & [11]\\
\dataset [ADS/Sa.CXO#Obs/08107] {NGC 4472}             & 12:29:47 & +08:00:02 & 16.4 & $0.115  \pm 0.005$  & $0.0914 \pm 0.0131$ & $0.53^{+0.16}_{-0.32}$ & 16.1  & A & [12]\\
\dataset [ADS/Sa.CXO#Obs/02072] {NGC 4552}             & 12:35:40 & +12:33:23 & 15.4 & $0.0398 \pm 0.0012$ & \nodata             & $0.53^{+0.12}_{-0.17}$ & 2.22  & A & [13]\\
\dataset [ADS/Sa.CXO#Obs/04415] {NGC 4636}             & 12:42:50 & +02:41:16 & 14.7 & $0.0281 \pm 0.0010$ & $0.0843 \pm 0.0005$ & $2.76^{+0.56}_{-0.91}$ & 37.6  & B & [14]\\
\dataset [ADS/Sa.CXO#Obs/03220] {NGC 4782}$^{\dagger}$ & 12:54:36 & -12:34:07 & 66.9 & $57.6   \pm 1.8$    & $66.8   \pm 6.7$    & $2.43^{+0.86}_{-1.43}$ & 13.9  & B & [15]\\
\dataset [ADS/Sa.CXO#Obs/09399] {NGC 5044}             & 13:15:24 & -16:23:08 & 31.4 & $0.0571 \pm 0.0018$ & $0.0126 \pm 0.0013$ & $4.18^{+1.18}_{-1.97}$ & 425   & B & [16]\\
\dataset [ADS/Sa.CXO#Obs/05907] {NGC 5813}             & 15:01:11 & +01:42:07 & 32.4 & $0.0260 \pm 0.0018$ & $0.334  \pm 0.054$  & $3.97^{+1.02}_{-2.36}$ & 180   & A & [17]\\
\dataset [ADS/Sa.CXO#Obs/07923] {NGC 5846}             & 15:06:29 & +01:36:20 & 24.9 & $0.0220 \pm 0.0014$ & $0.0609 \pm 0.0001$ & $0.88^{+0.30}_{-0.59}$ & 64.4  & B & [18]\\
\dataset [ADS/Sa.CXO#Obs/04972] {NGC 6269}             & 16:57:58 & +27:51:16 & 155  & $1.99   \pm 0.08$   & \nodata             & $1.54^{+0.49}_{-1.03}$ & 118   & C & [19]\\
\dataset [ADS/Sa.CXO#Obs/04194] {NGC 6338}             & 17:15:23 & +57:24:40 & 120  & $1.37   \pm 0.04$   & $0.451  \pm 0.045$  & $11.0^{+3.3}_{-6.9}$   & 388   & B & [20]\\
\dataset [ADS/Sa.CXO#Obs/02074] {NGC 7626}$^{\dagger}$ & 23:20:43 & +08:13:01 & 51.7 & $3.76   \pm 0.14$   & \nodata             & $0.39^{+0.09}_{-0.18}$ & 11.5  & B & [21]
\enddata
\tablecomments{
  Col. (1) Source name -- those with a dagger ($\dagger$) are poorly
  confined (see Section \ref{sec:jet}) and are not plotted in Figure
  \ref{fig:pcav};
  Col. (2) Right ascension;
  Col. (3) Declination;
  Col. (4) Luminosity distance;
  Col. (5) 1.4 GHz radio power;
  Col. (6) 200-400 MHz radio power;
  Col. (7) Cavity power calculated using buoyancy age;
  Col. (8) 0.3-2.0 keV X-ray luminosity within radius where $t_{\mathrm{cool}} \le 7.7$ Gyrs;
  Col. (9) Cavity system figure of merit (see Section \ref{sec:r&d}).
  Col. (10) References for other relevant studies:
  [1] \cite{2003ApJ...585..677P}, [2] \cite{2008ApJ...687..986D}, [3]
  \cite{2003MNRAS.343L..73W}, [4] \cite{2004ApJ...601..221K}, [5]
  \cite{2009ApJ...704.1586S}, [6] \cite{2010AAS...21545802L}, [7]
  \cite{2001ApJ...552..106B}, [8] \cite{2004ApJ...617..262S}, [9]
  \cite{2006ApJ...653..207D}, [10] \cite{2005ApJ...627..711Z}, [11]
  \cite{2010RAA....10..220X}, [12] \cite{2004ApJ...613..238B}, [13]
  \cite{2006ApJ...648..947M}, [14] \cite{2009ApJ...707.1034B}, [15]
  \cite{2007ApJ...664..804M}, [16] \cite{2009ApJ...705..624D}, [17]
  \cite{randall_n5813}, [18] \cite{2006MNRAS.372...21A}, [19]
  \cite{2009ApJ...694..479B}, [20] \cite{2007MNRAS.376..193J}, [21]
  \cite{2009ApJ...696.1431R},
}
\end{deluxetable}

%% file: figs.tex
\begin{center}
  \begin{figure}[htp]
    \begin{minipage}[htp]{0.5\linewidth}
      \includegraphics*[width=\textwidth, trim=30mm 5mm 40mm 15mm, clip]{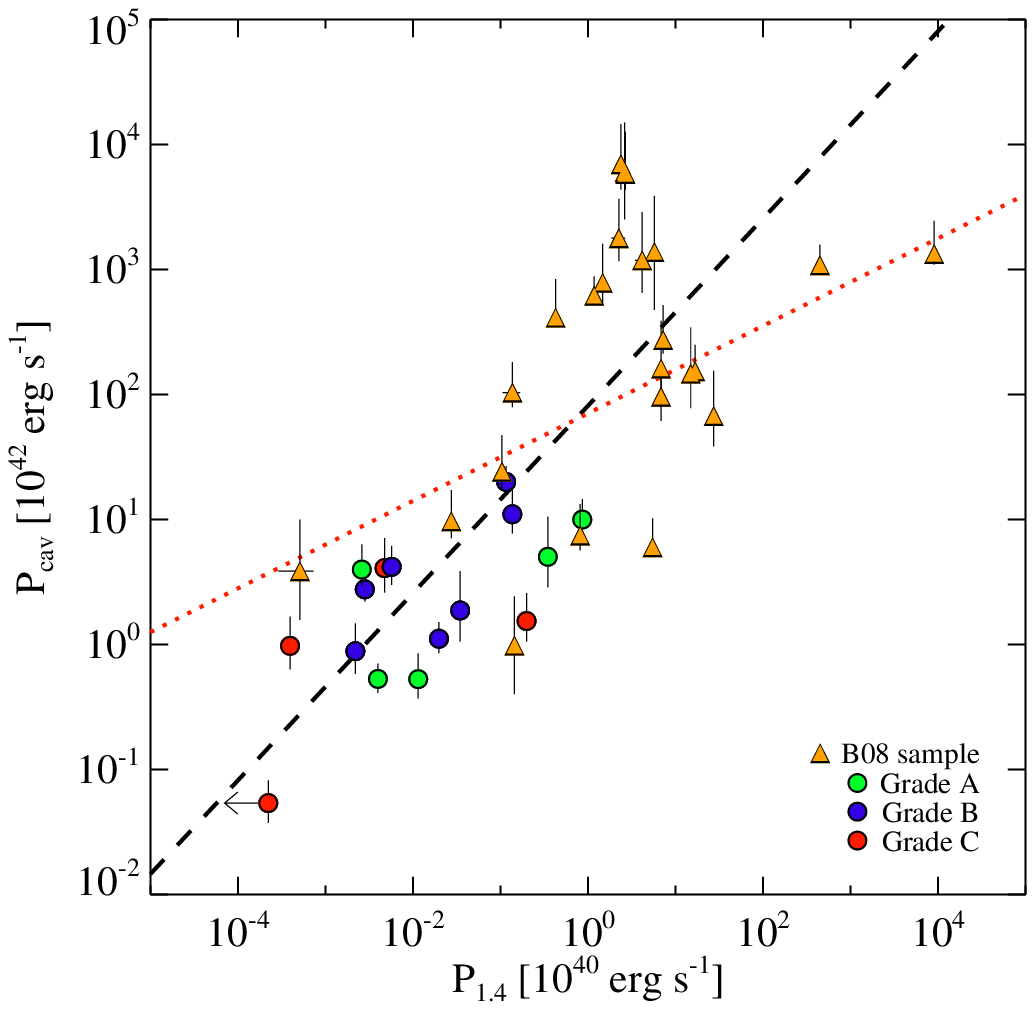}
    \end{minipage}
    \begin{minipage}[htp]{0.5\linewidth}
      \includegraphics*[width=\textwidth, trim=30mm 5mm 40mm 15mm, clip]{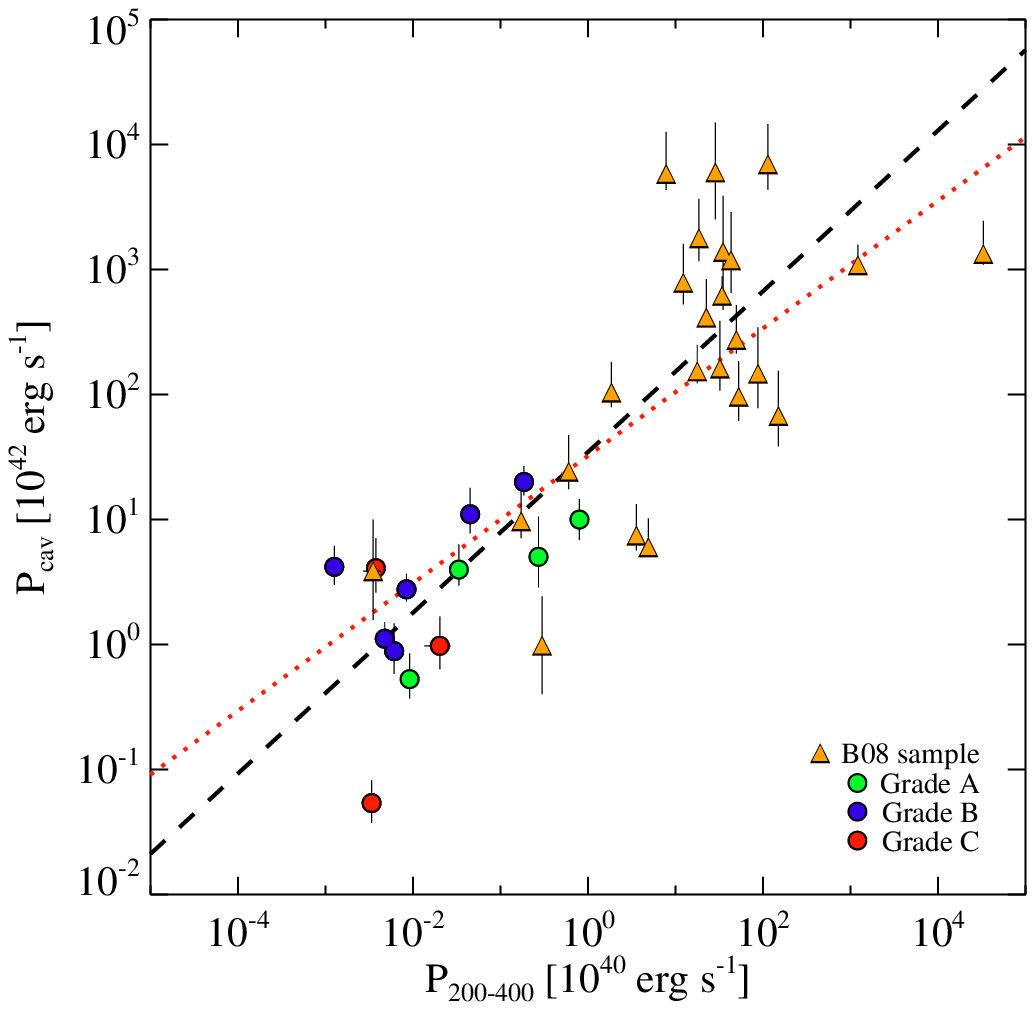}
    \end{minipage}
    \caption{Cavity power vs. radio power. Orange triangles represent
      the galaxy clusters and groups sample from B08. Filled circles
      represent our sample of gEs with colors representing the cavity
      system figure of merit (see Section \S\ref{sec:xray}): green =
      `A,' blue = `B,' and red = `C.' The dotted red lines represent
      the best-fit power-law relations presented in B08 using only the
      orange triangles. The dashed black lines represent our
      \bces\ best-fit power-law relations. {\it{Left:}} Cavity power
      vs. 1.4 GHz radio power. {\it{Right:}} Cavity power vs. 200-400
      MHz radio power.}
    \label{fig:pcav}
  \end{figure}
\end{center}

\begin{figure}[htp]
  \begin{center}
    \begin{minipage}[htp]{0.5\linewidth}
      \includegraphics*[width=\textwidth, trim=30mm 5mm 40mm 15mm, clip]{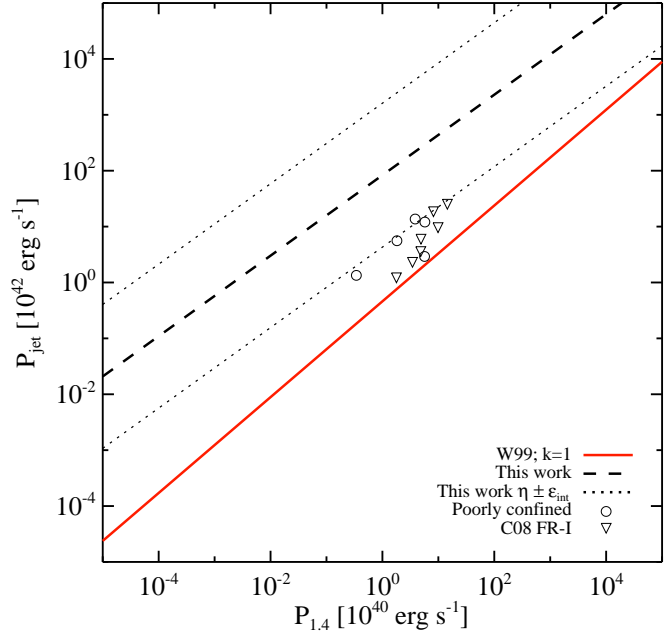}
      \caption{Comparison of scaling relations between jet power and
        radio luminosity. The solid red line represents the
        \citet[][W99]{w99} model with $k=1$. The dashed black line is
        our best-fit \pjet-\phigh\ relation (Equation
        \ref{eqn:high}). The dotted black lines denote the upper and
        lower limits of our best-fit relation after including
        intrinsic scatter of $\epsilon_{\mathrm{int}} = 1.3$ dex. The
        unfilled black circles denote the poorly confined sources
        discussed in Section \ref{sec:jet}, and the downfacing black
        triangles are FR-I sources taken from the sample in
        \citet[][C08]{2008MNRAS.386.1709C}.}
      \label{fig:radeff}
    \end{minipage}
  \end{center}
\end{figure}

\begin{figure}[htp]
  \begin{center}
    \begin{minipage}[htp]{\linewidth}
      \includegraphics*[width=\textwidth, trim=0mm 0mm 0mm 0mm, clip]{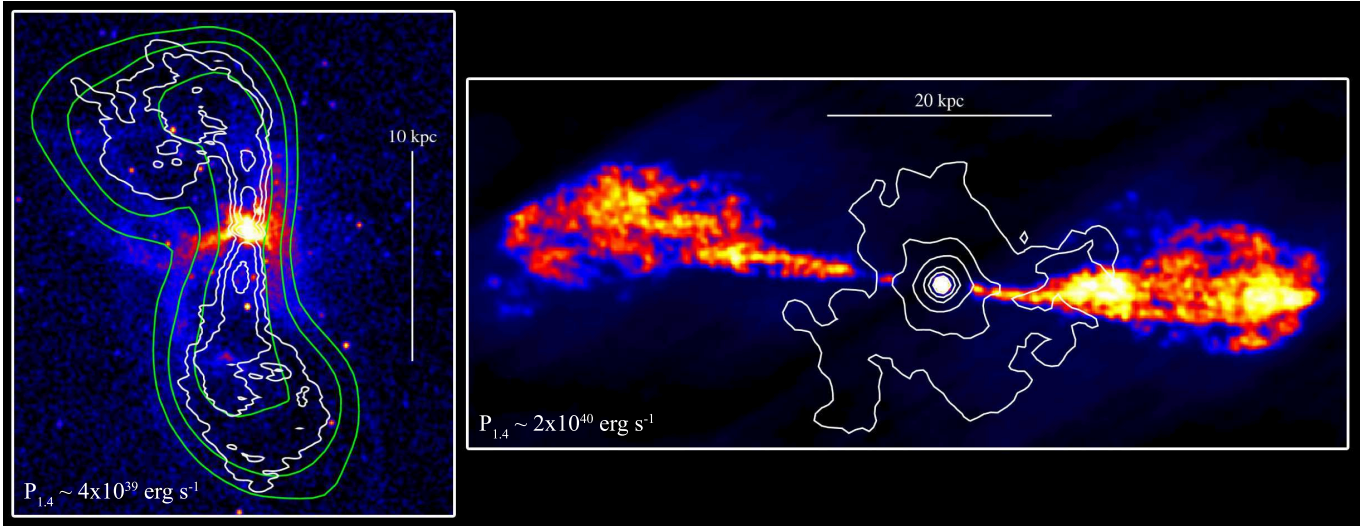}
      \caption{{\it{Left:}} \chandra\ X-ray image of the giant
        elliptical M84 (NGC 4374). Contours trace out 1.4 GHz radio
        emission as observed with VLA C-configuration (green) and
        AB-configuration (white) ranging from $\approx 0.5-50$ mJy in
        log spaced steps of 10 mJy. Note the displacement of the X-ray
        gas around the bipolar AGN jet outflows. M84 examplifies the
        typical interaction between an AGN outflow and a hot gaseous
        halo. {\it{Right:}} VLA B-configuration 1.4 GHz radio image of
        the AGN jets eminating from the giant elliptical NGC
        4261. White contours trace \chandra\ observed X-ray emission
        of the hot halo surrounding N4261. The contours cover the
        surface brightness range of $\approx 5-50$ cts arcsec$^{-2}$
        in linear spaced steps of 5 cts arcsec$^{-2}$. N4261
        demonstrates the characteristic traits of what we have termed
        poorly confined sources: a compact X-ray halo, small
        centralized cavities along the jets, FR-I-like radio
        morphology, and pluming beyond the ``edge'' of the X-ray
        halo.}
      \label{fig:pics}
    \end{minipage}
  \end{center}
\end{figure}

%% file: arxiv.bbl
\begin{thebibliography}{71}
\expandafter\ifx\csname natexlab\endcsname\relax\def\natexlab#1{#1}\fi

\bibitem[{{Akritas} \& {Bershady}(1996)}]{bces}
{Akritas}, M.~G., \& {Bershady}, M.~A. 1996, \apj, 470, 706

\bibitem[{{Allen} {et~al.}(2006){Allen}, {Dunn}, {Fabian}, {Taylor}, \&
  {Reynolds}}]{2006MNRAS.372...21A}
{Allen}, S.~W., {Dunn}, R.~J.~H., {Fabian}, A.~C., {Taylor}, G.~B., \&
  {Reynolds}, C.~S. 2006, \mnras, 372, 21

\bibitem[{{Baldi} {et~al.}(2009{\natexlab{a}}){Baldi}, {Forman}, {Jones},
  {Kraft}, {Nulsen}, {Churazov}, {David}, \&
  {Giacintucci}}]{2009ApJ...707.1034B}
{Baldi}, A., {Forman}, W., {Jones}, C., {Kraft}, R., {Nulsen}, P., {Churazov},
  E., {David}, L., \& {Giacintucci}, S. 2009{\natexlab{a}}, \apj, 707, 1034

\bibitem[{{Baldi} {et~al.}(2009{\natexlab{b}}){Baldi}, {Forman}, {Jones},
  {Nulsen}, {David}, {Kraft}, \& {Simionescu}}]{2009ApJ...694..479B}
{Baldi}, A., {Forman}, W., {Jones}, C., {Nulsen}, P., {David}, L., {Kraft}, R.,
  \& {Simionescu}, A. 2009{\natexlab{b}}, \apj, 694, 479

\bibitem[{{Best} {et~al.}(2007){Best}, {von der Linden}, {Kauffmann},
  {Heckman}, \& {Kaiser}}]{best07}
{Best}, P.~N., {von der Linden}, A., {Kauffmann}, G., {Heckman}, T.~M., \&
  {Kaiser}, C.~R. 2007, \mnras, 379, 894

\bibitem[{{Beuing} {et~al.}(1999){Beuing}, {Dobereiner}, {B{\"o}hringer}, \&
  {Bender}}]{1999MNRAS.302..209B}
{Beuing}, J., {Dobereiner}, S., {B{\"o}hringer}, H., \& {Bender}, R. 1999,
  \mnras, 302, 209

\bibitem[{{Biller} {et~al.}(2004){Biller}, {Jones}, {Forman}, {Kraft}, \&
  {Ensslin}}]{2004ApJ...613..238B}
{Biller}, B.~A., {Jones}, C., {Forman}, W.~R., {Kraft}, R., \& {Ensslin}, T.
  2004, \apj, 613, 238

\bibitem[{B\^{i}rzan {et~al.}(2010)B\^{i}rzan, McNamara, Nulsen, Carilli, , \&
  Wise}]{birzan08err}
B\^{i}rzan, L., McNamara, B.~R., Nulsen, P. E.~J., Carilli, C.~L., , \& Wise,
  M.~W. 2010, The Astrophysical Journal, 709, 546

\bibitem[{{B{\^\i}rzan} {et~al.}(2008){B{\^\i}rzan}, {McNamara}, {Nulsen},
  {Carilli}, \& {Wise}}]{birzan08}
{B{\^\i}rzan}, L., {McNamara}, B.~R., {Nulsen}, P.~E.~J., {Carilli}, C.~L., \&
  {Wise}, M.~W. 2008, \apj, 686, 859

\bibitem[{{B{\^\i}rzan} {et~al.}(2004){B{\^\i}rzan}, {Rafferty}, {McNamara},
  {Wise}, \& {Nulsen}}]{birzan04}
{B{\^\i}rzan}, L., {Rafferty}, D.~A., {McNamara}, B.~R., {Wise}, M.~W., \&
  {Nulsen}, P.~E.~J. 2004, \apj, 607, 800

\bibitem[{{Blandford} \& {Konigl}(1979)}]{1979ApJ...232...34B}
{Blandford}, R.~D., \& {Konigl}, A. 1979, \apj, 232, 34

\bibitem[{{Blanton} {et~al.}(2001){Blanton}, {Sarazin}, \&
  {Irwin}}]{2001ApJ...552..106B}
{Blanton}, E.~L., {Sarazin}, C.~L., \& {Irwin}, J.~A. 2001, \apj, 552, 106

\bibitem[{{Bock} {et~al.}(1999){Bock}, {Large}, \& {Sadler}}]{sumss1}
{Bock}, D.~C.-J., {Large}, M.~I., \& {Sadler}, E.~M. 1999, \aj, 117, 1578

\bibitem[{{Bower} {et~al.}(2006){Bower}, {Benson}, {Malbon}, {Helly}, {Frenk},
  {Baugh}, {Cole}, \& {Lacey}}]{bower06}
{Bower}, R.~G., {Benson}, A.~J., {Malbon}, R., {Helly}, J.~C., {Frenk}, C.~S.,
  {Baugh}, C.~M., {Cole}, S., \& {Lacey}, C.~G. 2006, \mnras, 370, 645

\bibitem[{{Br{\"u}ggen}(2003)}]{2003ApJ...592..839B}
{Br{\"u}ggen}, M. 2003, \apj, 592, 839

\bibitem[{{Cavagnolo} {et~al.}(2009){Cavagnolo}, {Donahue}, {Voit}, \&
  {Sun}}]{accept}
{Cavagnolo}, K.~W., {Donahue}, M., {Voit}, G.~M., \& {Sun}, M. 2009, \apjs,
  182, 12

\bibitem[{{Cavaliere} \& {Fusco-Femiano}(1978)}]{betamodel}
{Cavaliere}, A., \& {Fusco-Femiano}, R. 1978, \aap, 70, 677

\bibitem[{{Churazov} {et~al.}(2001){Churazov}, {Br{\"u}ggen}, {Kaiser},
  {B{\"o}hringer}, \& {Forman}}]{2001ApJ...554..261C}
{Churazov}, E., {Br{\"u}ggen}, M., {Kaiser}, C.~R., {B{\"o}hringer}, H., \&
  {Forman}, W. 2001, \apj, 554, 261

\bibitem[{{Condon}(1992)}]{1992ARA&A..30..575C}
{Condon}, J.~J. 1992, \araa, 30, 575

\bibitem[{{Condon} {et~al.}(1998){Condon}, {Cotton}, {Greisen}, {Yin},
  {Perley}, {Taylor}, \& {Broderick}}]{nvss}
{Condon}, J.~J., {Cotton}, W.~D., {Greisen}, E.~W., {Yin}, Q.~F., {Perley},
  R.~A., {Taylor}, G.~B., \& {Broderick}, J.~J. 1998, \aj, 115, 1693

\bibitem[{{Croston} {et~al.}(2008){Croston}, {Hardcastle}, {Birkinshaw},
  {Worrall}, \& {Laing}}]{2008MNRAS.386.1709C}
{Croston}, J.~H., {Hardcastle}, M.~J., {Birkinshaw}, M., {Worrall}, D.~M., \&
  {Laing}, R.~A. 2008, \mnras, 386, 1709

\bibitem[{{Croton} {et~al.}(2006){Croton}, {Springel}, {White}, {De Lucia},
  {Frenk}, {Gao}, {Jenkins}, {Kauffmann}, {Navarro}, \& {Yoshida}}]{croton06}
{Croton}, D.~J., {Springel}, V., {White}, S.~D.~M., {De Lucia}, G., {Frenk},
  C.~S., {Gao}, L., {Jenkins}, A., {Kauffmann}, G., {Navarro}, J.~F., \&
  {Yoshida}, N. 2006, \mnras, 365, 11

\bibitem[{{David} {et~al.}(2009){David}, {Jones}, {Forman}, {Nulsen},
  {Vrtilek}, {O'Sullivan}, {Giacintucci}, \&
  {Raychaudhury}}]{2009ApJ...705..624D}
{David}, L.~P., {Jones}, C., {Forman}, W., {Nulsen}, P., {Vrtilek}, J.,
  {O'Sullivan}, E., {Giacintucci}, S., \& {Raychaudhury}, S. 2009, \apj, 705,
  624

\bibitem[{{David} {et~al.}(2006){David}, {Jones}, {Forman}, {Vargas}, \&
  {Nulsen}}]{2006ApJ...653..207D}
{David}, L.~P., {Jones}, C., {Forman}, W., {Vargas}, I.~M., \& {Nulsen}, P.
  2006, \apj, 653, 207

\bibitem[{{De Young}(2006)}]{2006ApJ...648..200D}
{De Young}, D.~S. 2006, \apj, 648, 200

\bibitem[{{De Young} {et~al.}(2008){De Young}, {O'Neill}, \&
  {Jones}}]{2008ASPC..386..343D}
{De Young}, D.~S., {O'Neill}, S.~M., \& {Jones}, T.~W. 2008, in Astronomical
  Society of the Pacific Conference Series, Vol. 386, Extragalactic Jets:
  Theory and Observation from Radio to Gamma Ray, ed. {T.~A.~Rector \& D.~S.~De
  Young}, 343--+

\bibitem[{{Diehl} \& {Statler}(2008)}]{2008ApJ...687..986D}
{Diehl}, S., \& {Statler}, T.~S. 2008, \apj, 687, 986

\bibitem[{{Dunn} {et~al.}(2006){Dunn}, {Fabian}, \&
  {Celotti}}]{2006MNRAS.372.1741D}
{Dunn}, R.~J.~H., {Fabian}, A.~C., \& {Celotti}, A. 2006, \mnras, 372, 1741

\bibitem[{{Dunn} {et~al.}(2005){Dunn}, {Fabian}, \&
  {Taylor}}]{2005MNRAS.364.1343D}
{Dunn}, R.~J.~H., {Fabian}, A.~C., \& {Taylor}, G.~B. 2005, \mnras, 364, 1343

\bibitem[{{Fabian} {et~al.}(2000){Fabian}, {Sanders}, {Ettori}, {Taylor},
  {Allen}, {Crawford}, {Iwasawa}, {Johnstone}, \& {Ogle}}]{perseus1}
{Fabian}, A.~C., {Sanders}, J.~S., {Ettori}, S., {Taylor}, G.~B., {Allen},
  S.~W., {Crawford}, C.~S., {Iwasawa}, K., {Johnstone}, R.~M., \& {Ogle}, P.~M.
  2000, \mnras, 318, L65

\bibitem[{{Falcke} \& {Biermann}(1995)}]{1995A&A...293..665F}
{Falcke}, H., \& {Biermann}, P.~L. 1995, \aap, 293, 665

\bibitem[{{Falle}(1991)}]{1991MNRAS.250..581F}
{Falle}, S.~A.~E.~G. 1991, \mnras, 250, 581

\bibitem[{{Finoguenov} \& {Jones}(2001)}]{2001ApJ...547L.107F}
{Finoguenov}, A., \& {Jones}, C. 2001, \apjl, 547, L107

\bibitem[{{Finoguenov} {et~al.}(2008){Finoguenov}, {Ruszkowski}, {Jones},
  {Br{\"u}ggen}, {Vikhlinin}, \& {Mandel}}]{2008ApJ...686..911F}
{Finoguenov}, A., {Ruszkowski}, M., {Jones}, C., {Br{\"u}ggen}, M.,
  {Vikhlinin}, A., \& {Mandel}, E. 2008, \apj, 686, 911

\bibitem[{{Forman} {et~al.}(2007){Forman}, {Jones}, {Churazov}, {Markevitch},
  {Nulsen}, {Vikhlinin}, {Begelman}, {B{\"o}hringer}, {Eilek}, {Heinz},
  {Kraft}, {Owen}, \& {Pahre}}]{2007ApJ...665.1057F}
{Forman}, W., {Jones}, C., {Churazov}, E., {Markevitch}, M., {Nulsen}, P.,
  {Vikhlinin}, A., {Begelman}, M., {B{\"o}hringer}, H., {Eilek}, J., {Heinz},
  S., {Kraft}, R., {Owen}, F., \& {Pahre}, M. 2007, \apj, 665, 1057

\bibitem[{{Hart} {et~al.}(2009){Hart}, {Stocke}, \&
  {Hallman}}]{2009ApJ...705..854H}
{Hart}, Q.~N., {Stocke}, J.~T., \& {Hallman}, E.~J. 2009, \apj, 705, 854

\bibitem[{{Heinz} \& {Grimm}(2005)}]{2005ApJ...633..384H}
{Heinz}, S., \& {Grimm}, H.~J. 2005, \apj, 633, 384

\bibitem[{{Heinz} \& {Sunyaev}(2003)}]{2003MNRAS.343L..59H}
{Heinz}, S., \& {Sunyaev}, R.~A. 2003, \mnras, 343, L59

\bibitem[{{Jetha} {et~al.}(2007){Jetha}, {Ponman}, {Hardcastle}, \&
  {Croston}}]{2007MNRAS.376..193J}
{Jetha}, N.~N., {Ponman}, T.~J., {Hardcastle}, M.~J., \& {Croston}, J.~H. 2007,
  \mnras, 376, 193

\bibitem[{{Kauffmann} \& {Haehnelt}(2000)}]{2000MNRAS.311..576K}
{Kauffmann}, G., \& {Haehnelt}, M. 2000, \mnras, 311, 576

\bibitem[{{Kormendy} \& {Richstone}(1995)}]{1995ARA&A..33..581K}
{Kormendy}, J., \& {Richstone}, D. 1995, \araa, 33, 581

\bibitem[{{Kraft} {et~al.}(2004){Kraft}, {Forman}, {Churazov}, {Laslo},
  {Jones}, {Markevitch}, {Murray}, \& {Vikhlinin}}]{2004ApJ...601..221K}
{Kraft}, R.~P., {Forman}, W.~R., {Churazov}, E., {Laslo}, N., {Jones}, C.,
  {Markevitch}, M., {Murray}, S.~S., \& {Vikhlinin}, A. 2004, \apj, 601, 221

\bibitem[{{Kraft} {et~al.}(2003){Kraft}, {V{\'a}zquez}, {Forman}, {Jones},
  {Murray}, {Hardcastle}, {Worrall}, \& {Churazov}}]{2003ApJ...592..129K}
{Kraft}, R.~P., {V{\'a}zquez}, S.~E., {Forman}, W.~R., {Jones}, C., {Murray},
  S.~S., {Hardcastle}, M.~J., {Worrall}, D.~M., \& {Churazov}, E. 2003, \apj,
  592, 129

\bibitem[{{Lanz} {et~al.}(2010){Lanz}, {Jones}, {Forman}, {Ashby}, {Kraft}, \&
  {Hickox}}]{2010AAS...21545802L}
{Lanz}, L., {Jones}, C., {Forman}, W.~R., {Ashby}, M.~L.~N., {Kraft}, R., \&
  {Hickox}, R.~C. 2010, in Bulletin of the American Astronomical Society,
  Vol.~41, Bulletin of the American Astronomical Society, 483--+

\bibitem[{{Machacek} {et~al.}(2006){Machacek}, {Nulsen}, {Jones}, \&
  {Forman}}]{2006ApJ...648..947M}
{Machacek}, M., {Nulsen}, P.~E.~J., {Jones}, C., \& {Forman}, W.~R. 2006, \apj,
  648, 947

\bibitem[{{Machacek} {et~al.}(2007){Machacek}, {Kraft}, {Jones}, {Forman}, \&
  {Hardcastle}}]{2007ApJ...664..804M}
{Machacek}, M.~E., {Kraft}, R.~P., {Jones}, C., {Forman}, W.~R., \&
  {Hardcastle}, M.~J. 2007, \apj, 664, 804

\bibitem[{{Magliocchetti} \& {Br{\"u}ggen}(2007)}]{2007MNRAS.379..260M}
{Magliocchetti}, M., \& {Br{\"u}ggen}, M. 2007, \mnras, 379, 260

\bibitem[{{Magorrian} {et~al.}(1998){Magorrian}, {Tremaine}, {Richstone},
  {Bender}, {Bower}, {Dressler}, {Faber}, {Gebhardt}, {Green}, {Grillmair},
  {Kormendy}, \& {Lauer}}]{magorrian}
{Magorrian}, J., {Tremaine}, S., {Richstone}, D., {Bender}, R., {Bower}, G.,
  {Dressler}, A., {Faber}, S.~M., {Gebhardt}, K., {Green}, R., {Grillmair}, C.,
  {Kormendy}, J., \& {Lauer}, T. 1998, \aj, 115, 2285

\bibitem[{{Mathur} {et~al.}(2009){Mathur}, {Stoll}, {Krongold}, {Nicastro},
  {Brickhouse}, \& {Elvis}}]{2009arXiv0910.3691M}
{Mathur}, S., {Stoll}, R., {Krongold}, Y., {Nicastro}, F., {Brickhouse}, N., \&
  {Elvis}, M. 2009, ArXiv e-prints: 0910.3691

\bibitem[{{McNamara} {et~al.}(2000){McNamara}, {Wise}, {Nulsen}, {David},
  {Sarazin}, {Bautz}, {Markevitch}, {Vikhlinin}, {Forman}, {Jones}, \&
  {Harris}}]{2000ApJ...534L.135M}
{McNamara}, B.~R., {Wise}, M., {Nulsen}, P.~E.~J., {David}, L.~P., {Sarazin},
  C.~L., {Bautz}, M., {Markevitch}, M., {Vikhlinin}, A., {Forman}, W.~R.,
  {Jones}, C., \& {Harris}, D.~E. 2000, \apjl, 534, L135

\bibitem[{{Merloni} \& {Heinz}(2007)}]{2007MNRAS.381..589M}
{Merloni}, A., \& {Heinz}, S. 2007, \mnras, 381, 589

\bibitem[{{Miley}(1980)}]{1980ARA&A..18..165M}
{Miley}, G. 1980, \araa, 18, 165

\bibitem[{{Nulsen} {et~al.}(2005){Nulsen}, {Hambrick}, {McNamara}, {Rafferty},
  {B\^irzan}, {Wise}, \& {David}}]{herca}
{Nulsen}, P.~E.~J., {Hambrick}, D.~C., {McNamara}, B.~R., {Rafferty}, D.,
  {B\^irzan}, L., {Wise}, M.~W., \& {David}, L.~P. 2005, \apjl, 625, L9

\bibitem[{{O'Sullivan} {et~al.}(2003){O'Sullivan}, {Ponman}, \&
  {Collins}}]{2003MNRAS.340.1375O}
{O'Sullivan}, E., {Ponman}, T.~J., \& {Collins}, R.~S. 2003, \mnras, 340, 1375

\bibitem[{{Pellegrini} {et~al.}(2003){Pellegrini}, {Venturi}, {Comastri},
  {Fabbiano}, {Fiore}, {Vignali}, {Morganti}, \&
  {Trinchieri}}]{2003ApJ...585..677P}
{Pellegrini}, S., {Venturi}, T., {Comastri}, A., {Fabbiano}, G., {Fiore}, F.,
  {Vignali}, C., {Morganti}, R., \& {Trinchieri}, G. 2003, \apj, 585, 677

\bibitem[{{Pratt} {et~al.}(2009){Pratt}, {Croston}, {Arnaud}, \&
  {B{\"o}hringer}}]{2009A&A...498..361P}
{Pratt}, G.~W., {Croston}, J.~H., {Arnaud}, M., \& {B{\"o}hringer}, H. 2009,
  \aap, 498, 361

\bibitem[{{Puchwein} {et~al.}(2008){Puchwein}, {Sijacki}, \&
  {Springel}}]{2008ApJ...687L..53P}
{Puchwein}, E., {Sijacki}, D., \& {Springel}, V. 2008, \apjl, 687, L53

\bibitem[{{Rafferty} {et~al.}(2006){Rafferty}, {McNamara}, {Nulsen}, \&
  {Wise}}]{rafferty06}
{Rafferty}, D.~A., {McNamara}, B.~R., {Nulsen}, P.~E.~J., \& {Wise}, M.~W.
  2006, \apj, 652, 216

\bibitem[{{Randall} {et~al.}(2010){Randall}, {Forman}, {Giacintucci}, {Nulsen},
  {Sun}, {Jones}, {Churazov}, {David}, {Kraft}, {Donahue}, {Blanton},
  {Simionescu}, \& {Werner}}]{randall_n5813}
{Randall}, S.~W., {Forman}, W.~R., {Giacintucci}, S., {Nulsen}, P.~E.~J.,
  {Sun}, M., {Jones}, C., {Churazov}, E., {David}, L.~P., {Kraft}, R.,
  {Donahue}, M., {Blanton}, E.~L., {Simionescu}, A., \& {Werner}, N. 2010,
  ArXiv e-prints

\bibitem[{{Randall} {et~al.}(2009){Randall}, {Jones}, {Kraft}, {Forman}, \&
  {O'Sullivan}}]{2009ApJ...696.1431R}
{Randall}, S.~W., {Jones}, C., {Kraft}, R., {Forman}, W.~R., \& {O'Sullivan},
  E. 2009, \apj, 696, 1431

\bibitem[{{Silk} \& {Rees}(1998)}]{1998A&A...331L...1S}
{Silk}, J., \& {Rees}, M.~J. 1998, \aap, 331, L1

\bibitem[{{Sivakoff} {et~al.}(2004){Sivakoff}, {Sarazin}, \&
  {Carlin}}]{2004ApJ...617..262S}
{Sivakoff}, G.~R., {Sarazin}, C.~L., \& {Carlin}, J.~L. 2004, \apj, 617, 262

\bibitem[{{Sun}(2009)}]{2009ApJ...704.1586S}
{Sun}, M. 2009, \apj, 704, 1586

\bibitem[{{Verkhodanov} {et~al.}(1997){Verkhodanov}, {Trushkin}, {Andernach},
  \& {Chernenkov}}]{cats}
{Verkhodanov}, O.~V., {Trushkin}, S.~A., {Andernach}, H., \& {Chernenkov},
  V.~N. 1997, in Astronomical Society of the Pacific Conference Series, Vol.
  125, Astronomical Data Analysis Software and Systems VI, ed. G.~{Hunt} \&
  H.~{Payne}, 322--+

\bibitem[{{Weinmann} {et~al.}(2006){Weinmann}, {van den Bosch}, {Yang}, {Mo},
  {Croton}, \& {Moore}}]{2006MNRAS.372.1161W}
{Weinmann}, S.~M., {van den Bosch}, F.~C., {Yang}, X., {Mo}, H.~J., {Croton},
  D.~J., \& {Moore}, B. 2006, \mnras, 372, 1161

\bibitem[{{Whiteoak}(1970)}]{1970ApL.....5...29W}
{Whiteoak}, J.~B. 1970, \aplett, 5, 29

\bibitem[{{Willott} {et~al.}(1999){Willott}, {Rawlings}, {Blundell}, \&
  {Lacy}}]{w99}
{Willott}, C.~J., {Rawlings}, S., {Blundell}, K.~M., \& {Lacy}, M. 1999,
  \mnras, 309, 1017

\bibitem[{{Wise} {et~al.}(2007){Wise}, {McNamara}, {Nulsen}, {Houck}, \&
  {David}}]{hydraa}
{Wise}, M.~W., {McNamara}, B.~R., {Nulsen}, P.~E.~J., {Houck}, J.~C., \&
  {David}, L.~P. 2007, \apj, 659, 1153

\bibitem[{{Worrall} {et~al.}(2003){Worrall}, {Birkinshaw}, \&
  {Hardcastle}}]{2003MNRAS.343L..73W}
{Worrall}, D.~M., {Birkinshaw}, M., \& {Hardcastle}, M.~J. 2003, \mnras, 343,
  L73

\bibitem[{{Xu} {et~al.}(2010){Xu}, {Gu}, {Gu}, {Zhang}, {Wang}, \&
  {An}}]{2010RAA....10..220X}
{Xu}, H., {Gu}, J., {Gu}, L., {Zhang}, Z., {Wang}, Y., \& {An}, T. 2010,
  Research in Astronomy and Astrophysics, 10, 220

\bibitem[{{Zezas} {et~al.}(2005){Zezas}, {Birkinshaw}, {Worrall}, {Peters}, \&
  {Fabbiano}}]{2005ApJ...627..711Z}
{Zezas}, A., {Birkinshaw}, M., {Worrall}, D.~M., {Peters}, A., \& {Fabbiano},
  G. 2005, \apj, 627, 711

\end{thebibliography}
